\documentclass{PoS}

\title{Jet Correlations from PHENIX: Low-$\rm{p_{T}}$ to High-$\rm{p_{T}}$}
\ShortTitle{Jet Correlations from PHENIX: Low-$p_{T}$ to
High-$p_{T}$}

\author{\speaker{Nathan Grau}%
         \thanks{For the PHENIX Collaboration}\\
        Columbia University, Nevis Labs\\
        E-mail: \email{ncgrau@nevis.columbia.edu}}

\abstract{PHENIX has measured many different two-particle azimuthal
correlations in several different colliding systems, beam energies,
$p_{T}$ windows, etc. The initial striking results from the first
full energy Au+Au run have been confirmed, \textit{i.e.}~the
observation of away-side suppression at intermediate $p_{T}$ and the
shape modification at low $p_{T}$. With the new high-statistics data
these results have been extended to further explore the physics
resulting in these modified correlations. In this contribution we
present a wide variety of correlation results at high-$p_{T}$ ($>$ 5
GeV/c) and at low $p_{T}$ ($\sim$ 1-4 GeV/c). We discuss the
implication of these data on energy loss models and on the
possibilities of determining medium properties from these
correlations.}

\FullConference{The 2nd edition of the International Workshop --- Correlations and Fluctuations in Relativistic Nuclear Collisions --- \\
         July 7-9 2006\\
         Galileo Galilei Institute, Florence, Italy}

\begin{document}

\section{Introduction}
With the advent of RHIC, study of hard-scattering physics in
heavy-ion collisions has been possible. Hard-scattering physics
touches both the hard and soft physics that is measured at RHIC. The
jets that we are sensitive to, those with energies of a $\sim$5 to
20 GeV, typically fragment into relatively soft hadrons. As a result
direct reconstruction of these jets is very difficult because it is
virtually impossible to disentangle the jet fragmentation from those
from bulk particle production. Jet quenching scenarios generally
yield results that modify the bulk particle production, albeit at a
small level. The radiated energy from the hard-scattered parton
results in increased particle production, probably at low $p_{T}$.
Also, $v_2$ as a function of $p_{T}$ deviates from hydrodynamical
calculations near 1-2 GeV/c~\cite{PHENIXv2}. Further, jet quenching
at high-$p_{T}$ was proposed as a probe of the soft medium produced
in RHIC collisions.

Jet physics at RHIC has been accessed by single particle observables
such as high-$p_{T}$ spectra, but, more directly by two-particle
correlations. In two-particle correlations one triggers on a
hard-scattering process by selecting events with a high-$p_{T}$
particle (the trigger) and measures the azimuthal (and longitudinal)
distribution of other particles in the event (the associated
particles) with respect to the trigger. Experimentally two-particle
correlations are determined from a correlation function,
$C\left(\Delta\phi\right)$ by measuring the \textit{real}
distribution of pairs within the triggered events and removing
detector correlations by measuring the \textit{mixed} distribution
of pairs where a trigger is in one event and an associated particle
is in a second event. The resulting correlation function is usually
arbitrarily normalized since the shape of the distribution is most
directly measurable. However, with the knowledge of the associated
particle efficiency it is possible to determine the normalization
such that the resulting distribution is the yield of pairs per
trigger~\cite{JiaCorrNorm}. This is summarized as
\begin{equation}
\frac{1}{N_{trig}} \frac{dN}{d\Delta\phi} \propto
C\left(\Delta\phi\right) \propto
\frac{\rm{Real}\left(\Delta\phi\right)}{\rm{Mix}\left(\Delta\phi\right)}
\end{equation}
In the two-component model of azimuthal correlations in A+A
collisions, the shape is expected to be characterized by
\begin{eqnarray}
\frac{1}{N_{trig}} \frac{dN}{d\Delta\phi} & = &
\frac{N}{2\pi}\left(1 +
2v_{2}^{trigg}v_{2}^{assoc}\cos\left(2\Delta\phi\right)\right) +
\nonumber \\
& &
\frac{Y_{N}}{\sqrt{2\pi}\sigma_{N}}\exp{\left(\frac{-\Delta\phi^{2}}{2\sigma_{N}^{2}}\right)}
+ \nonumber \\
& &
\frac{Y_{F}}{\sqrt{2\pi}\sigma_{F}}\exp{\left(\frac{-\left(\Delta\phi-\pi\right)^{2}}{2\sigma_{F}^{2}}\right)}
\end{eqnarray}
The first term is the correlation from the elliptic flow. The second
term is the near-side correlation for two particles which fragment
from a single jet. The last term is the away-side jet which results
from the fragment of a second, recoil jet correlated with the
trigger jet. Here we have explicitly assumed the hard scattering is
the result of a 2$\rightarrow$2 process.

Studies of these correlations began early at RHIC and there were
some initial striking results. At intermediate $p_{T}$ with a
trigger hadrons of $\sim$ 4 GeV/c and associated hadrons $\sim$ 2
GeV/c, the recoil jet was almost completely
suppressed~\cite{STARdisappear}. This was not seen in p+p and d+Au
collisions~\cite{PPG039} where a ``normal'', approximately Gaussian,
away-side is present. The measurement of away-side suppression is
complimentary to the single particle suppression results and
indicated that the away-side is much more strongly affected by
energy loss than the near-side. By triggering on a high-$p_{T}$
particle, the measured correlations have a bias to jets produced
near the surface of the interaction region. Therefore, the near-side
jet has a short path length through the medium while the away-side
potentially traverses a substantial distance across the whole of the
interaction region. Measurements at lower $p_{T}$, near 2 GeV/c for
the trigger and the associated particle, an away-side shape was
measured but its shape was much broader and it was peaked away from
$\Delta\phi=\pi$~\cite{PPG032}. This result stirred much theoretical
interest into bending jets~\cite{BentJetsASW}\cite{BentJetsHwa},
mach cones~\cite{StoeckerMachShocks}\cite{JorgeMachCone}, and
Cerenkov radiation~\cite{CerenkovCone}.

In the recent high-statistics data samples from RHIC of Au+Au and
Cu+Cu, work has been done to extend these initial studies to further
understand the physics that has been hinted at. In this contribution
we review the recent work on two-particle correlations from the
PHENIX experiment both at low-$p_{T}$ and at high-$p_{T}$. At
high-$p_{T}$ it is interesting to look for the effects of jet
quenching, \textit{i.e.}~measurements of the away-side jet
broadening which should accompany the yield suppression from jet
quenching. At low-$p_{T}$ we would like to study the structure that
was measured and attempt to rule out different scenarios which could
provide an explanation for it, thereby possibly making contact with
properties of the medium.

The organization is as follows. The first section deals with the
high-$p_{T}$ correlations and their results. We discuss the
implications of these data on energy loss scenarios. In the next
section lower-$p_{T}$ correlations are discussed in some detail
where we argue that existing data begin to distinguish between
different models of the away-side jet structure. We end with a
summary of the landscape of two-particle correlations as it exists
at this point.

\section{High-$p_{T}$ Triggered Correlations}
The high-statistics Au+Au and Cu+Cu data sets taken at RHIC in 2004
and 2005 are an improvement on the previous data in two respects.
First, it is possible to extend the $p_{T}$ reach of the initial
two-particle results. Second, since the jet signal-to-background is
only a few percent, with the higher statistics it is possible to
observe a significant signal above the background where it was not
possible with the initial data set. These two aspects of the data
will be exploited in the next two sections.

\begin{figure}[h!tb]
\includegraphics[width=0.5\textwidth]{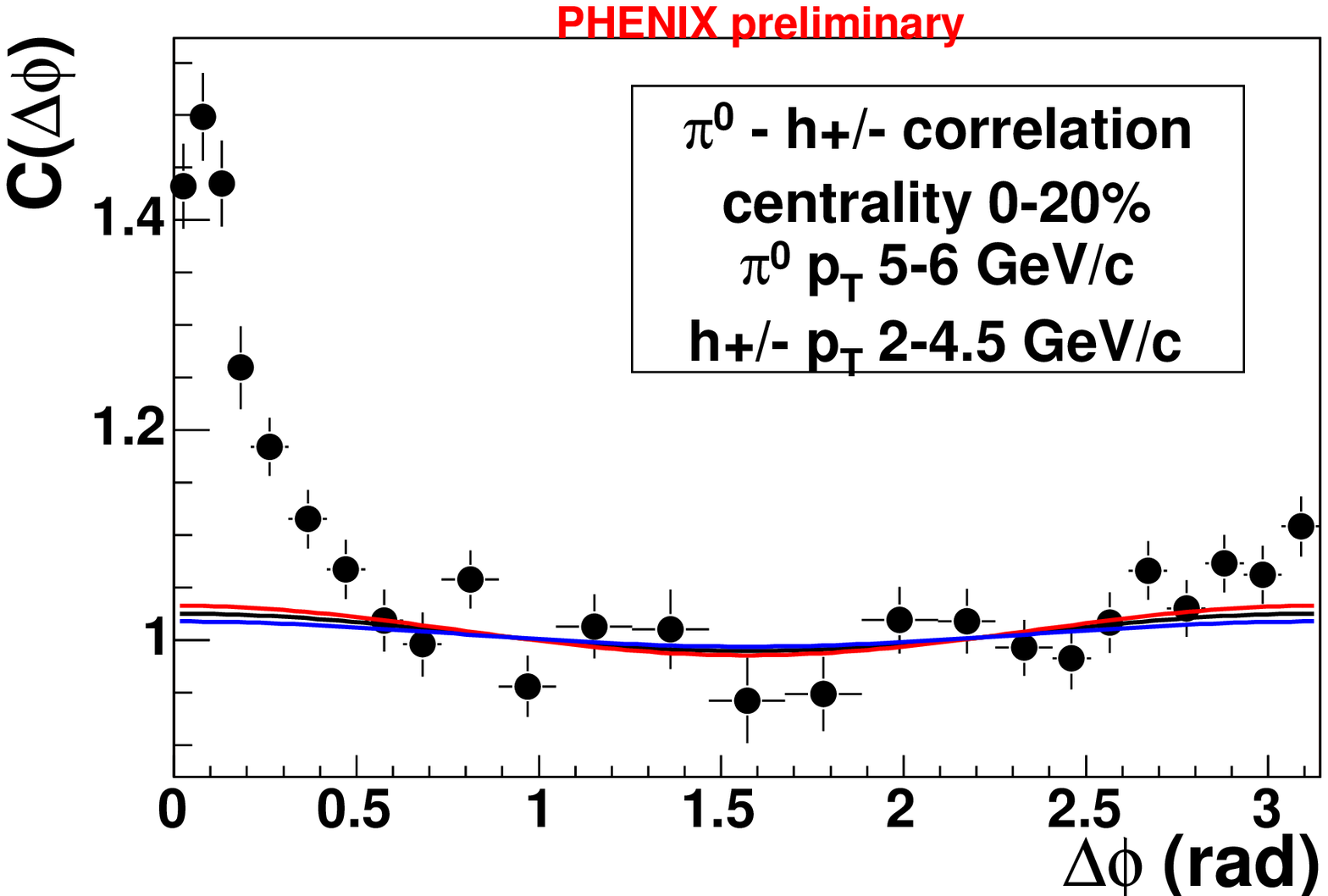}
\includegraphics[width=0.5\textwidth]{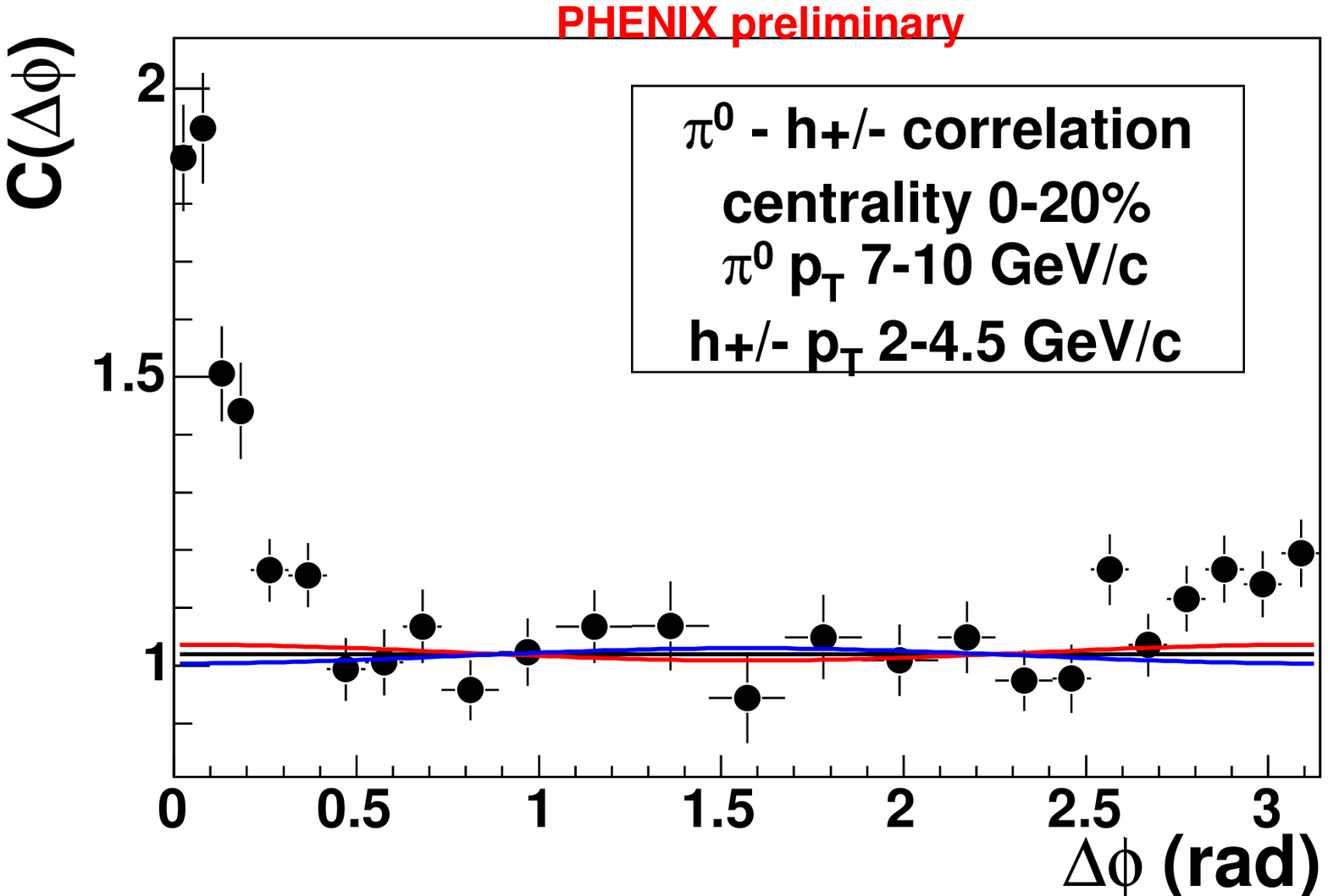}
\caption{High-$p_{T}$ $\pi^{0}$-h correlations in 0-20\% central
Au+Au collisions. The trigger $\pi^{0}$ has a $p_{T}$ of 5-6 GeV/c
(left) and 7-10 GeV/c (right) and the associated hadron has a
$p_{T}$ of 2.5-4.0 GeV/c. The lines indicated the contribution of
the $v_{2}$ correlation.}\label{fig:AuAuPi0hHighPt}
\end{figure}

In order to begin energy loss studies from two-particle correlations
and because of the surface bias of these correlations, it is
important to measure the away-side jet in the Au+Au environment.
Fig.~\ref{fig:AuAuPi0hHighPt} shows trigger $\pi^{0}$ correlations
with associated hadrons in 0-20\% Au+Au collisions. The $\pi^{0}$
triggers are above 5 GeV/c and the associated hadrons are 2-4.5
GeV/c. These $C\left(\Delta\phi\right)$ are normalized such that the
background around which the flow modulates is approximately unity.
In this case the strength of the correlation signal is the
signal-to-background of the jets. A clear near-side jet peak is seen
above the background. What had not been seen before this data is the
existence of excess yield above the $v_{2}$ correlation in the
away-side. This is the away-side jet and we can measure a
statistically significant yield for triggers of 5 GeV/c and
associated hadrons of 2 GeV/c.

Since the away-side is measurable with the high statistics in Au+Au,
we can move to study the energy loss of the away-side. The
expectation from radiative energy loss models is that the away-side
jet should not only be suppressed but it also should be broadened.
BDMPS showed that one component of the broadening is due to multiple
scattering the parton undergoes as it
radiates~\cite{BDMPSBroadening}. It was also recently shown that the
radiated gluons have a broad angular spectrum and, when they
subsequently hadronize and are measured as associated particles,
they should significantly contribute to the expected
broadening~\cite{IvanBroadening}.

\begin{figure}[h!tb]
\includegraphics[width=1.0\textwidth]{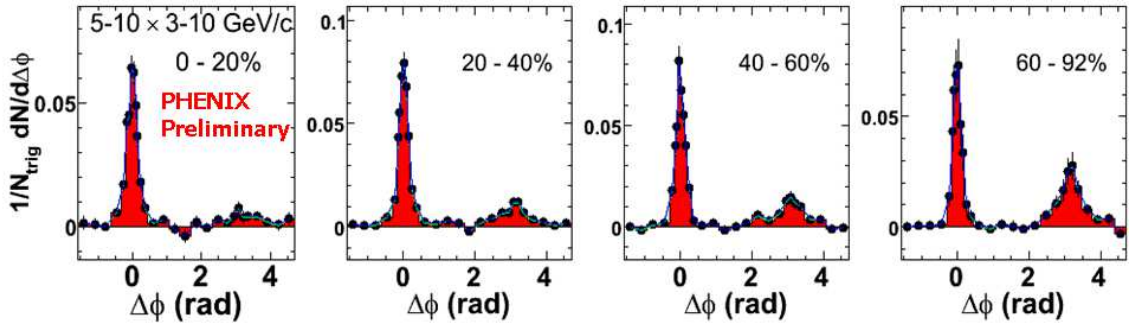}
\caption{High-$p_{T}$ h-h correlations for different Au+Au
centrality classes, left to right, 0-20\%, 20-40\%, 40-60\%, and
60-92\%. The trigger hadron has a $p_{T}$ of 5-10 GeV/c and the
associated hadron has a $p_{T}$ of 3-10 GeV/c. These are measured
pair per trigger distributions after the $v_{2}$ correlation has
been subtracted.}\label{fig:AuAuhhHighPt}
\end{figure}

A measurement of the yields at high-$p_{T}$ confirms the initial
result that the away-side yield is suppressed. This is seen in
Fig.~\ref{fig:AuAuhhHighPt} which plots several different h-h
correlations in several Au+Au centrality classes and for a constant
trigger and associated range. The near side yield changes very
little as a function of centrality. This is contrasted by the
away-side yield which visibly decreases from most peripheral to most
central collisions. This away-side suppression is qualitatively
consistent with the recent results from
STAR~\cite{STARARun4AwayJet}.

\begin{figure} [t!b]
\begin{center}
\includegraphics[width=0.5\textwidth]{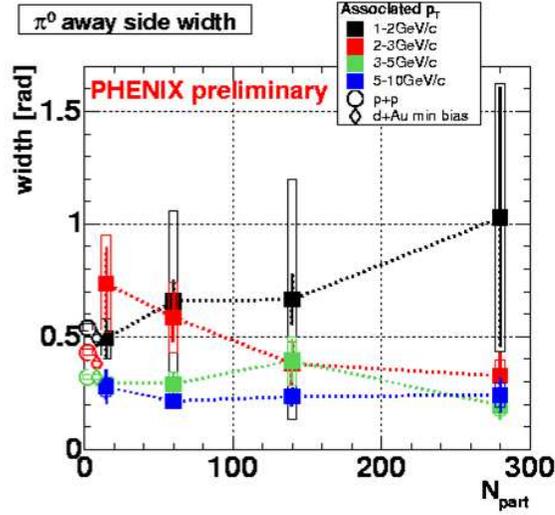}
\end{center}
\caption{Away-side Gaussian widths from 5-10 GeV/c trigger
$\pi^{0}$-h correlations as a function of centrality for different
associated hadron $p_{T}$ ranges. The lowest two curves correspond
to the 3-5 GeV/c and 5-10 GeV/c hadron ranges where the away-side
jet is most significant. The left most open points correspond to the
p+p and d+Au results.}\label{fig:AuAuPi0hWidths}
\end{figure}

A corresponding broadening should accompany the suppressed yield
that is measured. Such a measurement has been made by PHENIX in
$\pi^{0}$-h correlations and is shown in
Fig.~\ref{fig:AuAuPi0hWidths}. This figure plots the centrality
dependence of the away-side Gaussian width for 5-10 GeV/c triggered
$\pi^{0}$-h correlations. If one focuses on the lowest two curves
corresponding to the 3-5 GeV/c and 5-10 GeV/c associated hadron
ranges where the jet signal is most significant, the width is
consistent with the p+p values and flat as a function of centrality.
No significant broadening is measured. This is also in qualitative
agreement with similar measurements from
STAR~\cite{STARARun4AwayJet}.

The question from the data is how can the away-side jet be
suppressed but not broadened by energy loss? It was argued before
this data became available that the two-particle correlations result
from tangentially emitted jets. That is, they are biased toward the
surface and both near and away jets traverse little
medium~\cite{TangentialJets}. A suppression results from a
surface-to-volume suppression of the jet cross-section. The lack of
broadening results from the lack of interaction in the medium. This
would imply that there is a black interior where jets which recoil
into the center are absorbed and that the observed suppression is a
geometrical effect. This unfortunately does little to aid in the
understanding of the produced medium other than it is very opaque.

Another possibility is that the jets to which the two-particle
correlations are measuring are those sensitive to
$p_{0}\delta\left(\Delta E\right)$, the probability not to lose
energy~\cite{RenkPuncthrough}. In their Monte Carlo, which includes
expansion, the authors reproduce the suppression pattern of the
published STAR data. Further, they find that although there is a
surface bias, a non-negligible fraction of the di-jets are produced
near the center of the collision zone. The suppression results again
from the value of $p_{0}\delta\left(\Delta E\right)$. The lack of
broadening results from lack of energy loss.

This $p_{0}\delta\left(\Delta E\right)$ term is implicit in the
Gyulassy-Levai-Vitev (GLV) and explicit in the
Armesto-Salgado-Wiedemann (ASW) energy loss
models~\cite{ASWQuenchingWeights}. This term physically results from
the Poisson fluctuations in the number of scattering centers.
According to Ref.~\cite{ASWQuenchingWeights}
$p_{0}\delta\left(\Delta E\right)$ could be around 30\% from quark
jets and 10\% from gluon jets for multiple soft scattering. For
single hard scattering this probability is only slightly smaller.

It seems that both explanations of suppression but lack of
significant broadening result from a very opaque medium where the
two-particle correlation carry little information from the interior.
It is possible that the only information accessible by the
high-$p_{T}$ correlations is simply the probability not to interact.

\section{Low-$p_{T}$ Triggered Correlations}
In contrast to high-$p_{T}$ correlations, significant modification
of low-$p_{T}$ correlations has been observed. In the initial 200
GeV Au+Au collisions the away-side correlations are broadened such
that there is significant correlations beyond
$\Delta\phi=\pi\pm\pi/2$ and that the shape is peaked away from
$\pi$. The high-statistics data set allows further study of this
effect.

\begin{figure}[h!tb]
\begin{center}
\includegraphics[width=0.6\textwidth]{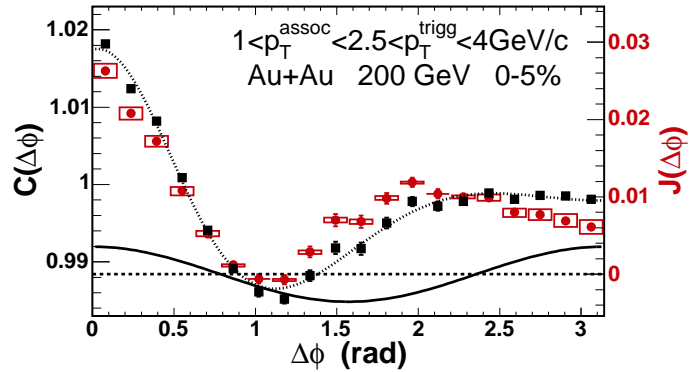}
\end{center}
\caption{Example of a low-$p_{T}$ h-h correlation in 0-5\% central
Au+Au collisions for trigger hadrons from 2.5-4.0 GeV/c and
associated hadrons from 1-2.5 GeV/c. The points with the dashed line
fit is the correlation function, the solid line indicates the
$v_{2}$ correlation, and the (red) boxed points indicate the
resulting jet shape after background subtraction.
}\label{fig:AuAuhhLowPt}
\end{figure}

The left panel of Fig.~\ref{fig:AuAuhhLowPt} shows an example of a
low-$p_{T}$ correlation function prior to and after the $v_{2}$
background subtraction. The measured correlation function (points
with the dashed line) is flat on the away-side. Therefore, any
harmonic function subtracted from the correlation function will
result in a minimum at $\Delta\phi=\pi$. This is what is observed
after the background subtraction.

\begin{figure}[h!tb]
\begin{center}
\includegraphics[width=0.8\textwidth]{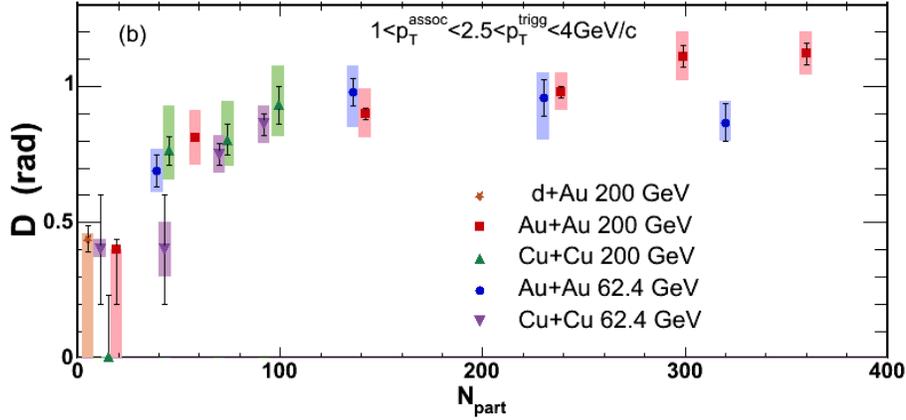}
\end{center}
\caption{The $D$ parameter (see text) from h-h correlations with
triggers from 2.5-4.0 and associated hadrons from 1.0-2.5 GeV/c as a
function of centrality for Au+Au and Cu+Cu collisions both at 200
GeV and 62.4 GeV.}\label{fig:AuAuSplittingParameter}
\end{figure}

To quantify this structure PHENIX has assumed the away-side shape is
a double gaussian symmetrically distributed around $\Delta\phi=\pi$,
that is
\begin{equation}
C_{away}\left(\Delta\phi\right) \propto
\exp{\left(\frac{-\left(\Delta\phi-\pi-D\right)^{2}}{2w^2}\right)} +
\exp{\left(\frac{-\left(\Delta\phi-\pi+D\right)^{2}}{2w^2}\right)}
\end{equation}
with the parameter $D$ quantifying the peak position of the
away-side. This parameter is plotted in
Fig.~\ref{fig:AuAuSplittingParameter} as a function of centrality
for Au+Au and Cu+Cu collisions both at 200 and 62.4 GeV. What is
observed is that the data follow, within errors, a smooth trend as a
function of $N_{part}$. This could indicate that the structure is
determined by a property of the medium instead of, for example, the
energy density. It is interesting to note recent studies from CERES
indicate that there is a qualitatively similar structure at SPS
energies~\cite{SPSCFs}.

There are many theoretical interpretations of the structure that was
formed. It was pointed out initially, before the data were
available, that jets could propagate through the medium with a
velocity above the speed of sound and produce Mach
shocks~\cite{StoeckerMachShocks}. One prediction of these models is
that the Mach angle is independent of the jet velocity,
\textit{i.e.}~the particle $p_{T}$.

One competing model to Mach cones is conical emission due to
Cerenkov radiation~\cite{CerenkovCone}. This model requires a set of
bound states in the matter produced resulting in an index of
refraction of the medium. The given index of refraction is
determined by the bound state spectrum. Because it is Cerenkov
radiation the cone angle will depend on the jet energy or particle
$p_{T}$. The $p_{T}$ dependence of the different models was proposed
as one possible way to determine if the conical emission was due to
Cerenkov radiation or Mach shocks.

The final set of models that are competing with the conical emission
models are bent jet scenarios~\cite{BentJetsASW}\cite{BentJetsHwa}.
From the statistical nature of two-particle correlations, it is not
clear that the two peak structure exists because of events having
alternating bent jets or from a single event having conical
emission. Two examples of bending jets are 1) longitudinal flow
distorting the jet~\cite{BentJetsASW} and 2) partons multiple
scattering away from the dense central medium~\cite{BentJetsHwa}.

Experimental measurements have been made in order to test these
different scenarios. Since the workshop, data has appeared from
PHENIX on the $p_{T}$ dependence of the two-particle correlations
shown in Fig.~\ref{fig:AuAuhhLowPt}. These correlations indicate
that there is little to no $p_{T}$ dependence to the $D$
parameter~\cite{PaulsRecentPPG}. At the least this would indicate
that the simple bound states spectrum from Ref.~\cite{CerenkovCone}
is not correct. More strongly, this data disfavors the Cerenkov cone
scenario. Still the bent jet scenarios cannot be explicitly ruled
out.

\begin{figure}[h!tb]
\begin{center}
\includegraphics[width=0.7\textwidth]{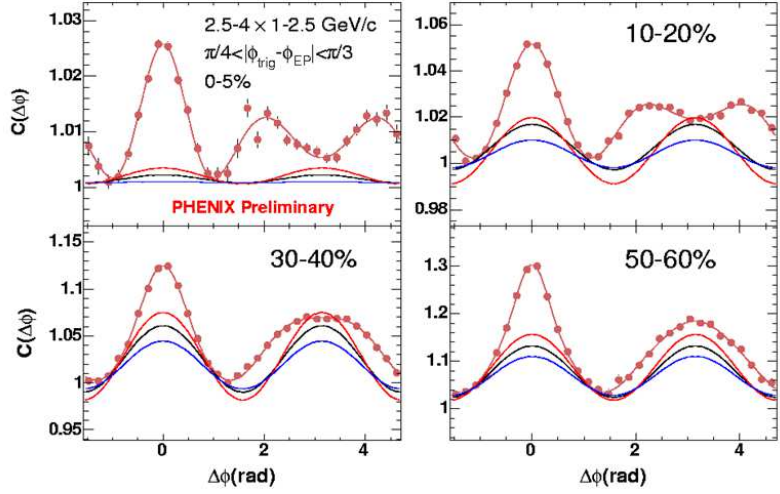}
\end{center}
\caption{Centrality dependence of h-h correlation functions prior to
$v_{2}$ subtraction where the trigger is fixed to be $\pi/4 <
|\phi_{trig} - \phi_{RP}| < \pi/3$ with respect to the reaction
plane (RP). The trigger range is 2.5-4.0 GeV/c and the associated
range is 1.0-2.5 GeV/c. The lines indicate the $v_{2}$ correlation.
In this reaction plane bin $v_{2}^{trig}$ has its smallest absolute
value.}\label{fig:AuAuhhLowPtSmallV2}
\end{figure}

Another interesting avenue of study is the reaction plane dependence
of these low-$p_{T}$ correlations. The reaction plane dependence
gives another handle on the $v_{2}$ systematics. The reaction plane
dependent correlations have a trigger in a particular bin in
$|\phi_{trig}-\phi_{EP}|$ (EP for event plane). In this case the
$v_{2}^{trig}$ is not equal to the inclusive $v_{2}^{trig}$ but
varies between bins in a well defined way~\cite{RPDepOfV2}. In fact,
$v_{2}^{trig}$ changes signs from positive when the trigger is along
the reaction plane to negative when the trigger is perpendicular to
the reaction plane. Depending on the reaction plane resolution there
is a bin in which the $v_{2}^{trig}$ is nearly zero.
Fig.~\ref{fig:AuAuhhLowPtSmallV2} shows different h-h correlations
for a fixed 2.5-4.0 GeV/c trigger and 1.0-2.5 GeV/c associated and a
fixed $\pi/4 < |\phi_{trig} - \phi_{RP}| < \pi/3$ reaction plane
bin. In this reaction plane bin $v_{2}^{trig}$ has its smallest
absolute value. What is observed in the most central two bins is
that the double-peaked structure exists prior to the background
subtraction.

\begin{figure}[t!b]
\begin{center}
\includegraphics[width=0.7\textwidth]{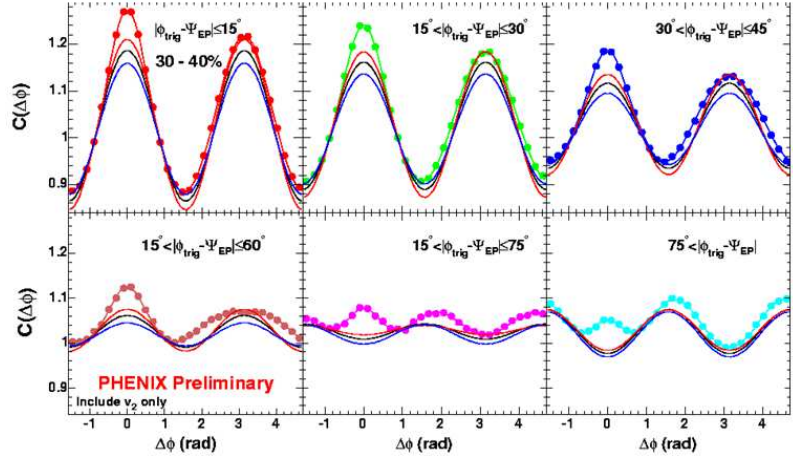}
\end{center}
\caption{Reaction plane dependence of 2.5-4.0 GeV/c trigger and
1.0-2.5 GeV/c associated h-h correlations in 30-40\% central Au+Au
collisions. The lines indicate the $v_{2}$
correlation.}\label{fig:AuAuhhLowPtRPDep}
\end{figure}

\begin{figure}[h!tb]
\begin{center}
\includegraphics[width=0.6\textwidth]{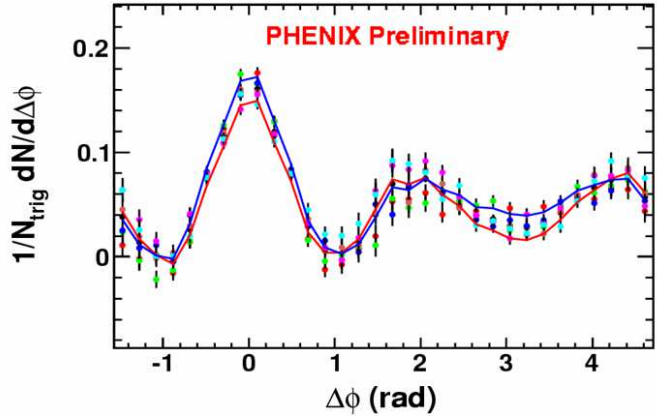}
\end{center}
\caption{Compilation of the $v_{2}$-subtracted pair per trigger
distribution for all reaction-plane dependent
correlations.}\label{fig:AuAuLowPtPerTrigger}
\end{figure}

The reaction plane dependence can be explored for which each have
different $v_{2}^{trig}$ systematics. These are shown in
Fig.~\ref{fig:AuAuhhLowPtRPDep} where each of the six reaction plane
bins in the 30-40\% central Au+Au collisions for the same trigger
and associated combination as in Fig.~\ref{fig:AuAuhhLowPtSmallV2}
are plotted. It is clear that the $v_{2}^{trig}$ contribution
changes quite dramatically between each of the reaction plane bins.
The $v_{2}$-subtracted correlations are plotted together as the per
trigger yields in Fig.~\ref{fig:AuAuLowPtPerTrigger}. The
observation is that the structure is, within errors, independent of
the trigger orientation with respect to the reaction plane. It is
important to stress that this result is after subtracting very
different $v_{2}$ values for each of the different reaction plane
bins. Such a measurement requires good control over the $v_{2}$
systematics. The physics implication of this may be an indication
that the data does not support bent jets from a flowing medium since
the peak angle would be dependent on the trigger's orientation with
respect to the reaction plane. Still, other bent jet scenarios
cannot be explicitly ruled out.

\section{Summary and Conclusions}
Two-particle correlations are still a developing field with many
results but no clear, consistent picture. At high-$p_{T}$ the data
from both STAR and PHENIX indicate a strongly suppressed yield on
the away-side. However, the required corresponding strong broadening
of the away-side based on radiative energy loss models is not
observed. The implications of this is under investigation. The
possibilities could be that, if the central core is true ``black'',
the correlations are only possibly sensitive to tangential jets or
to the non-interacting jets.

At lower trigger $p_{T}$, the away-side is strongly modified both in
the yield and in the shape. This shape seems to be independent of
the system size and collision species and with the trigger
orientation with respect to the reaction plane. The lack of $p_{T}$
dependence of the peak position of the away-side disfavors models of
Cerenkov radiation. The independence of the peak position on the
trigger orientation with respect to the reaction plane also
disfavors bent jets due to a flowing medium. However, other bent jet
scenarios have not be explicitly ruled out.

\end{document}